\documentclass[conference]{IEEEtran}
\IEEEoverridecommandlockouts
\usepackage{cite}
\usepackage{amsmath,amssymb,amsfonts}
\usepackage{algorithmic}
\usepackage{graphicx}
\usepackage{textcomp}
\usepackage{xcolor}
\usepackage{url}
\def\BibTeX{{\rm B\kern-.05em{\sc i\kern-.025em b}\kern-.08em
    T\kern-.1667em\lower.7ex\hbox{E}\kern-.125emX}}

\begin{document}

\title{Security Analysis of a Communication Protocol: MQTT\\
{\footnotesize \textsuperscript{}}
}

\author{\IEEEauthorblockN{Clarisse Sousa}
\textit{Polytechnic of Porto - School of Engineering}\\
\IEEEauthorblockA{\textit{Informatics Engineering} \\
Porto, Portugal \\
0009-0001-1793-2328}
\and

\IEEEauthorblockN{Ricardo Venâncio}
\textit{Polytechnic of Porto - School of Engineering}\\
\IEEEauthorblockA{\textit{Informatics Engineering} \\
Porto, Portugal \\
0009-0009-2697-9898}
\and

\IEEEauthorblockN{Luís Ribeiro}
\textit{Polytechnic of Porto - School of Engineering}\\
\IEEEauthorblockA{\textit{Informatics Engineering} \\
Porto, Portugal \\
0009-0002-7094-9978}
\and

\IEEEauthorblockN{Filipe Duarte}
\textit{Polytechnic of Porto - School of Engineering}\\
\IEEEauthorblockA{\textit{Informatics Engineering} \\
Porto, Portugal \\
0009-0004-0245-6826}
}

\maketitle

\begin{abstract}
This paper analyzes the security of the Message Queuing Telemetry Transport (MQTT) protocol in the context of the Internet of Things (IoT). The main objective consists of identifying vulnerabilities and proposing security improvements. Adopting a hybrid methodology, a theoretical review was combined with an experimental demonstration in a simulated Smart Home environment. Eavesdropping, Tampering, Denial of Service (DoS), and Brute Force attacks were executed and analyzed. The results evidenced critical risks due to the absence of robust encryption and authentication. Finally, mitigation strategies and best practices are proposed to strengthen MQTT implementations.
\end{abstract}


\section{Introduction}
Nowadays, most everyday devices are connected to the internet due to the production of cheap chips and high-bandwidth telecommunications. This allows millions of devices to have their data collected through sensors, processed, and used to generate intelligent actions \cite{ref1_aws_iot}. This concept is named the Internet of Things (IoT). Examples of IoT devices include sensors, cameras, lightbulbs, and connected vehicles. These devices can be used in applications such as smart homes, smart cities, and industrial environments.

With this inevitable adoption by millions of users worldwide, a new attack surface emerges, as malicious actors focus on the valuable assets of these users \cite{ref2_emnify}. Due to their nature, these devices share common characteristics that lead to severe threats \cite{ref3_fortinet_iot}:
\begin{itemize}
    \item \textbf{Limited Hardware:} IoT devices are often meant to be low-energy, meaning they are very simple and lack robust resources. This raises security issues, as the devices are unable to employ sophisticated methods against cyberattacks.
    \item \textbf{Diverse Communication Technologies:} In a distributed system, devices may use varying communication technologies depending on their needs (e.g., Wi-Fi, Bluetooth, Zigbee, LoRaWAN). Each technology has its own security mechanisms and vulnerabilities, making it complex to secure such a vast system entirely.
    \item \textbf{Vulnerable Components:} Due to their simplistic components, these devices are regularly susceptible to attacks.
\end{itemize}

Given that these devices operate under severe power and processing restrictions, it is crucial that the communication protocols used are lightweight and efficient. In this context, the Message Queuing Telemetry Transport (MQTT) protocol stands out. Designed for Machine-to-Machine (M2M) communication, it is optimized for limited bandwidth networks and low-resource devices \cite{ref11_mohanan}.

The main objective of this work is to perform a security analysis of the MQTT protocol to identify vulnerabilities and propose improvements. This includes identifying known and potential vulnerabilities through theoretical review, practically demonstrating exploits in a controlled environment, and proposing mitigation strategies and best practices.

\section{MQTT Protocol: Fundamental Concepts}
The emergence of distributed systems highlighted the need for more flexible data-sharing mechanisms, as the traditional client/server paradigm proved too rigid for dynamic and heterogeneous networks \cite{ref6_ajayi, ref7_nast}. 

\subsection{The Publisher/Subscriber Paradigm}
To address the limitations of the client/server model, the Publisher/Subscriber paradigm was introduced \cite{ref5_eugster}. In this model, nodes no longer depend directly on each other; instead, they rely on middleware to route the data they are interested in. A producer generates and publishes data, while a subscriber consumes only the data relevant to its interests. This mechanism of mediation ensures effective decoupling, where producers and consumers interact without requiring simultaneous connections or mutual knowledge. Depending on the system, the routing can be content-based, type-based, or topic-based \cite{ref8_shen}.

\subsection{Architecture and Operation}
MQTT, introduced in 1999, is a centralized, topic-based protocol \cite{ref12_elbasioni}. It relies on a central server called a \textit{broker} to manage topics, subscribers, and publishers. Key operational mechanisms include:
\begin{itemize}
    \item \textbf{Topics and Wildcards:} Topics function as channels for messages and follow a hierarchical structure (e.g., \texttt{/temperature/room}). Subscriptions can utilize wildcards: ``\texttt{+}'' substitutes exactly one level, while ``\texttt{\#}'' encompasses all subsequent sub-levels.
    \item \textbf{Quality of Service (QoS):} MQTT offers three levels of delivery guarantees. QoS 0 (\textit{At Most Once}) is a fire-and-forget approach. QoS 1 (\textit{At Least Once}) guarantees delivery but may result in duplicates. QoS 2 (\textit{Exactly Once}) ensures no duplicates but introduces higher latency.
    \item \textbf{Retained Messages and Last Will:} Publishers can mark a message to be retained by the broker, sending it automatically to new subscribers. The \textit{Last Will} mechanism allows a client to configure a message that the broker will publish automatically if the client disconnects abruptly.
    \item \textbf{Session Persistence:} Clients can connect with a \textit{clean session} flag. If disabled, the broker maintains persistent session information, delivering pending QoS 1 and QoS 2 messages when the client reconnects.
\end{itemize}

\subsection{Implementations and Use Cases}
MQTT represents a protocol specification rather than a single implementation. Widely used brokers include Eclipse Mosquitto, EMQX, HiveMQ, and VerneMQ \cite{ref20_cresswell, ref21_dallinger}. Client-side libraries, such as Eclipse Paho, exist for multiple programming languages. Due to its lightweight architecture, MQTT is heavily utilized in smart homes (domotics), industrial settings (exchanging data between PLCs and sensors), and smart city infrastructures (traffic and environmental monitoring).

\section{Security Mechanisms}
MQTT was designed to be lightweight, resulting in relatively limited native security mechanisms. Communication protection is generally ensured across external layers \cite{ref4_hivemq}.

\subsection{Authentication and Authorization}
The protocol supports authentication via Client Identifiers, X.509 Certificates, or standard username and password credentials. However, without external encryption, usernames and passwords are transmitted in plain text. Authorization is typically managed by the broker using Access Control Lists (ACLs), which define the topics and operations (publish/subscribe) permitted for each client.

\subsection{Transport Layer Security}
To mitigate plain-text transmission vulnerabilities, SSL/TLS is heavily recommended. TLS establishes a secure communication channel through an asymmetric cryptographic handshake, followed by symmetric encryption for continuous data transmission. While highly effective, TLS introduces significant computational overhead, which may not be viable for extremely resource-constrained devices.

\section{Vulnerability Analysis}
By default, MQTT transmits messages in plain text without encryption. This allows attackers to easily intercept sensitive data. Furthermore, brokers with default configurations often lack ACLs, allowing anonymous connections. 

\subsection{Known Vulnerabilities and Exploits}
A large-scale analysis identified approximately 425,000 public MQTT backends, of which 59\% allowed direct connections without any trust relationship \cite{ref13_tagliaro}. Many allowed wildcard subscriptions, granting attackers passive access to telemetry, location, health, and firmware update data. Furthermore, 99.84\% of analyzed MQTT/XMPP backends used insecure transport methods.

Several Common Vulnerabilities and Exposures (CVEs) have also been reported. For instance, CVE-2019-9749 exposes Fluent bit brokers to Distributed Denial of Service (DDoS) attacks \cite{ref14_cve2019}, while CVE-2018-19417 highlights a stack-based overflow in Contiki-NG servers allowing remote code execution \cite{ref15_cve2018}. Modern attack models have also demonstrated Man-in-the-Middle (MiTM) capabilities using adversarial machine learning to evade anomaly detection \cite{ref16_wong, ref17_chen, ref18_laghari}.

\subsection{Risk Matrix Assessment}
Based on the protocol's design and current deployment practices, several key threats emerge:
\begin{itemize}
    \item \textbf{Unauthorized Access (Critical Risk):} Permissive anonymous connections and missing ACLs enable wide-scale data harvesting facilitated by topic wildcards.
    \item \textbf{Denial of Service (Critical Risk):} Brokers are vulnerable to malformed packet parsing and act as a single point of failure in standard MQTT deployments.
    \item \textbf{Eavesdropping (High Risk):} Highly probable due to default plain-text transport and low TLS adoption rates among IoT developers.
    \item \textbf{Data Manipulation / Tampering (High Risk):} The absence of strong application-layer authentication allows active remote attackers to intercept and alter control messages.
\end{itemize}

\section{Experimental Environment and Demonstration}
A simulated Smart Home scenario was created to test these vulnerabilities. The environment was executed locally using Docker, containing distinct containers for communication, visualization, simulation, and attack scripts.

\begin{figure}[htbp]
\centerline{\includegraphics[width=\columnwidth]{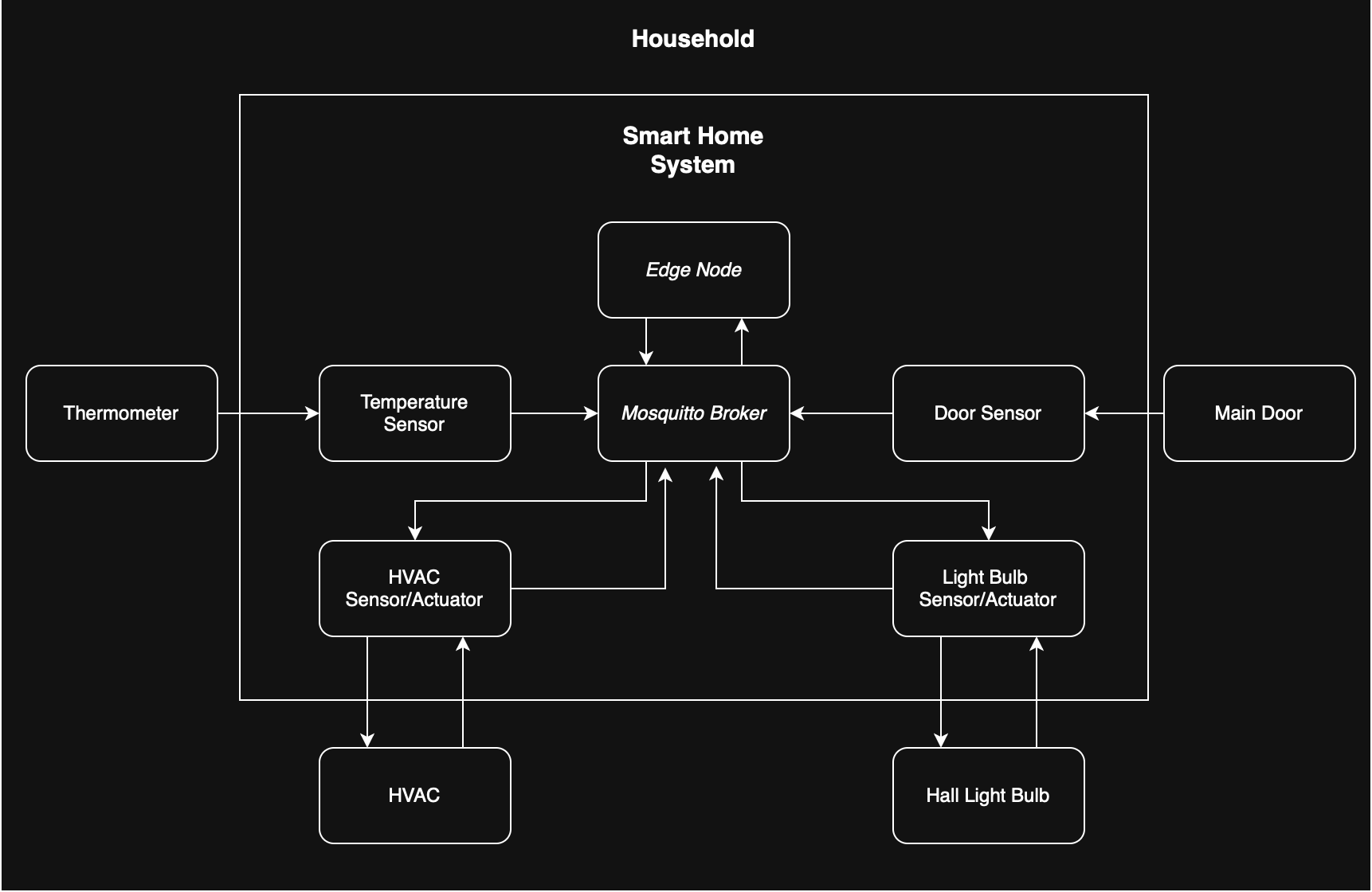}}
\caption{Architecture diagram of the simulated Smart Home environment.}
\label{fig:normal_arch}
\end{figure}

The core communication relied on the Eclipse Mosquitto broker (port 1883), chosen for its lightweight nature and maturity in edge scenarios \cite{ref20_cresswell, ref21_dallinger}. A Node-RED instance (port 1880) provided the visualization dashboard. As illustrated in Fig. \ref{fig:normal_arch}, an \textit{Edge Node} acted as the operation's brain, triggering an Air Conditioner (AC) if the temperature exceeded 24°C, and turning on a light if a door sensor registered as open.

\subsection{Eavesdropping}
The scenario was configured without TLS, meaning data traversed the network in plain text. An attacker on the same Docker network used a custom Python script, connected as an anonymous client, and subscribed to the \texttt{\#} wildcard. The script successfully captured sensitive plain-text payloads, including \texttt{\{"temperature": 23.4\}} and \texttt{\{"door\_state": "open"\}}. By logging this data into a CSV file, an attacker could study behavioral trends to determine when the house is empty.

\subsection{Tampering}
\begin{figure}[htbp]
\centerline{\includegraphics[width=\columnwidth]{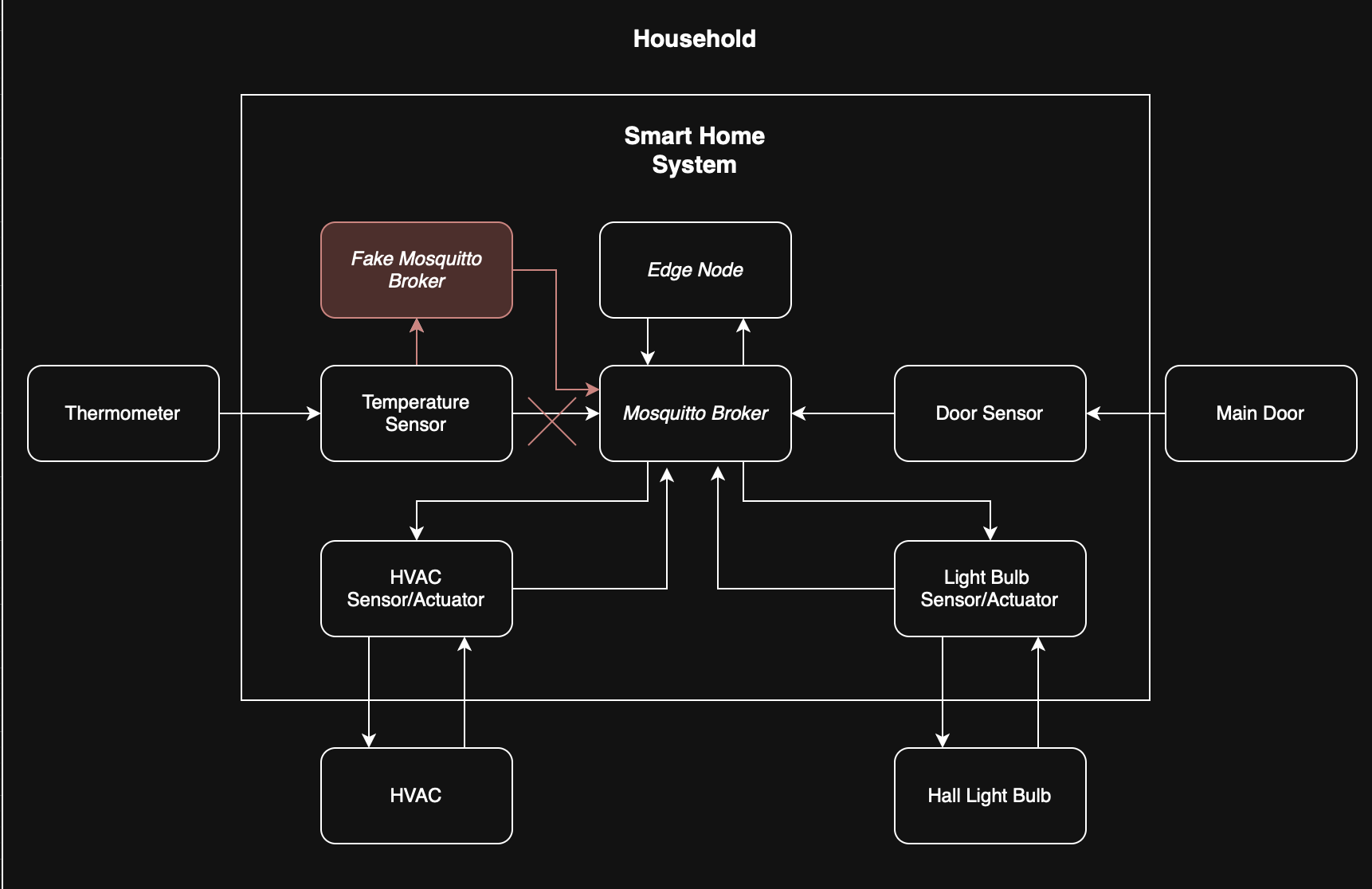}}
\caption{Diagram illustrating the Man-in-the-Middle (MiTM) tampering attack via a fake Mosquitto broker.}
\label{fig:tamper_arch}
\end{figure}

This attack utilized ARP Spoofing to execute a Man-in-the-Middle (MiTM) intercept. Using the \texttt{bettercap} tool within a Kali Linux container, the attacker flooded the network with Gratuitous ARP messages \cite{ref31_fortinet_arp, ref32_viescinski}. The network was tricked into routing the temperature sensor's traffic to a malicious proxy listening on port 1883, as shown in Fig. \ref{fig:tamper_arch}. 

The proxy adulterated the payload while carefully maintaining the original byte size to avoid corrupting the MQTT Remaining Length header. Consequently, the Node-RED dashboard displayed a tampered temperature of 999.9°C instead of the legitimate 24.56°C.

\subsection{Denial of Service}
To test availability, an open-source tool named \textit{MQTT Stresser} \cite{ref19_inovex} was deployed. The tool connected 3,000 concurrent clients, attempting to publish 100,000 messages with QoS 1. While the Mosquitto broker did not crash, service was severely degraded. A custom Python script evaluating end-to-end latency showed that message delays spiked from an average of 6 to 8 ms to over 60 seconds during the attack, effectively denying service to legitimate smart home components. Table \ref{tab:dos_latency} illustrates this severe service degradation by showing the recorded latency immediately before and during the attack.

\begin{table}[htbp]
\caption{End-to-End Message Latency Before and During DoS Attack}
\label{tab:dos_latency}
\begin{center}
\begin{tabular}{|c|c|c|}
\hline
\textbf{Message Seq.} & \textbf{Network State} & \textbf{Recorded Latency (s)} \\
\hline
1 & Normal & 0.007 \\
5 & Normal & 0.008 \\
10 & Normal & 0.011 \\
\hline
11 & DoS Initiated & 67.347 \\
12 & DoS Active & 66.350 \\
15 & DoS Active & 63.351 \\
20 & DoS Active & 58.350 \\
24 & DoS Active & 54.348 \\
\hline
\end{tabular}
\end{center}
\end{table}

\subsection{Credential Brute Force}
The Mosquitto broker does not natively limit login attempts or enforce password complexity. An attacker script systematically attempted authentication combinations. Searching a 36-character space (lowercase letters and numbers), a 4-character password (\texttt{1234}) was cracked in roughly 3,760 seconds (about one hour). Because the time requirement grows exponentially, a 5-character password would take an estimated 22 days. Furthermore, a timing attack was attempted, but Mosquitto's consistent response times prevented the extraction of sensitive temporal data.

\section{Mitigation and Best Practices}

\subsection{Eavesdropping and Tampering Mitigation}
Data must be protected via encryption. TLS ensures data confidentiality and integrity, preventing network-level eavesdropping and MiTM attacks \cite{ref22_alharbi}. However, for highly constrained devices where TLS overhead is too heavy, application-layer encryption should be used \cite{ref33_alkhafajee}. 

Recent studies propose utilizing Physical Unclonable Functions (PUFs) to generate unpredictable, dynamic symmetric keys without permanent storage \cite{ref23_gong}. Alternatively, lightweight software-oriented cryptographic algorithms like PRIDE and LEA offer high efficiency with low ROM/RAM consumption \cite{ref25_suryateja}. To ensure integrity without TLS, Message Authentication Codes (MACs) or Hash-based MACs (HMACs) must be appended to application payloads \cite{ref24_hintaw, ref34_gao}.

\subsection{Denial of Service Mitigation}
Application-layer DoS can be mitigated by enforcing strict authentication and ACL privileges. Brokers like Mosquitto should also be configured with strict resource limits, such as \texttt{max\_packet\_size}, \texttt{message\_size\_limit}, and \texttt{max\_inflight\_bytes} \cite{ref26_morelli}. 

At the network layer, firewalls, IP whitelisting, and Network Access Control (NAC) technologies should be implemented \cite{ref27_mt}. Intrusion Detection Systems utilizing Signature-based Detection \cite{ref28_bronte} or Machine Learning models (such as Random Forest or Support Vector Machines) can accurately identify and filter malicious MQTT traffic \cite{ref29_dikii}.

\subsection{Brute Force Mitigation}
Strict password policies (minimum length, complex characters) must be enforced. To counter the lack of native rate-limiting, external security tools like \textit{Fail2ban} can be used to analyze log files and temporarily block IP addresses exhibiting suspicious, repeated login failures \cite{ref30_fail2ban}.

\section{Conclusion}
In its default form, MQTT lacks robust security mechanisms, leaving it highly vulnerable to a variety of exploits, particularly in environments with constrained resources. The simulated smart home scenario successfully demonstrated severe operational risks, validating vulnerabilities regarding broker availability (DoS), data integrity (Tampering), confidentiality (Eavesdropping), and weak authentication (Brute Force). All proposed attacks succeeded in compromising the system.

While comprehensive mitigations (such as transport and application-layer encryption, robust ACLs, and active network monitoring) were proposed in detail theoretically, it is important to note a limitation of this work: these defensive measures were not practically implemented and tested against the simulated attacks in the lab environment. 

Given the results observed in this analysis, it would be highly relevant for future work to conduct similar practical security evaluations on other communication protocols frequently utilized in edge and IoT contexts, such as the Advanced Message Queuing Protocol (AMQP), Constrained Application Protocol (CoAP), Data Distribution Service (DDS), and Apache Kafka.

\bibliographystyle{IEEEtran}
\bibliography{references}

\end{document}